**Spectroscopic assessment of charge mobility in organic semiconductors**

*Andrey Yu. Sosorev[1,2]\*, Dmitry R. Maslennikov[1,2], Elizaveta V. Feldman[1], Ivan Yu.*

*Chernyshov,[3,4] Vladimir V. Bruevich[1,2], George G. Abashev[5,6], Oleg V. Borshchev[7], Sergei A.*

*Ponomarenko[7], Mikhail V. Vener[3,4], and Dmitry Yu. Paraschuk[1]\**

[1]Faculty of Physics and International Laser Center, M.V. Lomonosov Moscow State University, Leninskie Gory 1, Moscow 119991, Russia

[2]Institute of Spectroscopy of the Russian Academy of Sciences, Fizicheskaya Str., 5, Troitsk, Moscow 108840, Russia

[3]Department of Quantum Chemistry, Mendeleev University of Chemical Technology, Miusskaya Square 9, Moscow 125047, Russia

[4]Kurnakov Institute of General and Inorganic Chemistry, Russian Academy of Sciences, Leninskii prosp. 31, Moscow 119991, Russia

[5]Institute of Technical Chemistry Ural Branch of Russian Academy of Sciences, Ac. Koroleva st., 3, Perm 614013, Russia

[6]Perm State University Chemical Faculty, Bukireva st., 15, Perm 614990, Russia

[7]Enikolopov Institute of Synthetic Polymeric Materials, Russian Academy of Science, Profsoyuznaya 70, Moscow 117393, Russia

AUTHOR INFORMATION

**Corresponding Authors**

*E-mail: sosorev@physics.msu.ru (A.Yu.S.), paras@physics.msu.ru (D.Yu.P.)

**Abstract**. Rapid progress in organic electronics demands new highly efficient organic semiconducting materials. Nevertheless, only few materials have been created so far that show reliable band-like transport with high charge mobilities, which reflects the two main obstacles in the field: the poor understanding of charge transport in organic semiconductors (OSs) and the difficulty of its quantification in devices. Here, we present a spectroscopic method for assessment of the charge transport in organic semiconductors. We show that the intensities of the low-frequency Raman spectrum allow calculation of the dynamic disorder that limits the charge carrier mobility. The spectroscopically evaluated mobility clearly correlates with the device charge mobility reported for various OSs. The proposed spectroscopic method can serve as a powerful tool for a focused search of new materials and highlights the disorder bottleneck in the intrinsic charge transport in high-mobility organic semiconductors.

## 1. Introduction

High charge mobility in the active layers of various organic electronic devices is a prerequisite for their efficient operation. However, organic semiconductors (OSs) generally show charge mobilities, $\mu$, much below those of inorganic ones, and only a few OSs show reproducible $\mu$ exceeding that of amorphous silicon ($\mu \sim 1$ cm$^2$V$^{-1}$s$^{-1}$) — a workhorse of modern thin-film electronics.[1,2] Moreover, reliable $\mu$ measurements in OSs are difficult and time-consuming. For example, the commonly used method of $\mu$ measurement in organic field effect transistors (OFETs) is complicated by many factors such as the contacts, architecture, dielectric, etc., which results in various artifacts and pitfalls.[3] Therefore, the focused search for high-mobility OSs among huge number of candidates needs an effective approach for estimation of $\mu$ prior its measurements in devices.

OS consists of molecules bound by weak non-covalent interactions. Coherent charge transport resulting in high $\mu$ requires delocalization of a charge carrier over several molecules.[1,2,4] Such charge delocalization is promoted by considerable electronic coupling between the molecules, i.e., by large charge transfer integrals, $J$.[4] However, intra- and intermolecular vibrations (phonons) tend to localize the charge carrier at one molecule. This prevents coherent charge transport and results in charge hopping with low $\mu$, which is observed in most of OSs.[4,5] Interaction of charge carriers with vibrations — electron-phonon coupling — involves local (Holstein) and non-local (Peierls) contributions. The local one stems from modulation of molecular energy levels, mainly by intramolecular vibrations in the high-frequency (HF) spectral range, and is quantified by the reorganization energy, $\lambda$.[4] The non-local contribution results from modulation of $J$ mainly by low-frequency (LF) vibrations (inter-, intramolecular, or mixed[6,7]), and is characterized by the lattice distortion energy, $L$.[1] The frontier of LF range is usually set at $\omega = 200$ cm$^{-1}$, which corresponds to the energy of room-temperature thermal vibrations.

According to the recent theoretical studies,[8-11] the coupling of charge carriers to LF vibrations sets the limit for coherent charge transport in high-$\mu$ OSs. These vibrations are thermally populated at ambient conditions and hence have large atomic displacements with the result of strong dynamic disorder – variance of charge transfer integrals, $\sigma_J^2 = \left\langle \delta J^2 \right\rangle$.[4,12] Dynamic disorder decreases the delocalization length of charge carrier, $L_D$, and hence decreases $\mu$.[8-10] While multiple theoretical studies have addressed the impact of dynamic disorder on charge transport,[1,8-14] the experimental data that could verify the models and provide a criterion for screening promising OSs with weak electron-phonon interaction are extremely scarce.[15-19] Specifically, although the experimental LF vibrational spectra were used for verification of simulations,[19] direct probing of the dynamic disorder was not performed so far.

Raman spectroscopy — a standard tool for vibrational spectrum characterization of OSs[20-24] — gives the most direct access to electron-phonon coupling. The Raman signal is determined by the vibrational modulation of the electronic properties, namely, the material polarizability.[20,25] The latter strongly depends on charge delocalization, and high Raman signals in OSs are observed for vibrations that significantly affect the intra- and intermolecular π-electron conjugation and hence largely contribute to electron-phonon coupling.[6,26-28] Raman spectroscopy was previously applied for charge transport investigations in charge-transfer complexes. In those studies, analysis of the (pre-) resonant Raman intensities in the HF and LF ranges was exploited to extract information about local[29,30] and non-local[31,32] electron-phonon coupling, respectively. Although earlier attempts to associate the Raman intensities to the contributions of intramolecular modes to local electron-phonon interaction were made for other OSs,[33-36] further extension of this approach to address the dynamic disorder and estimate $\mu$ in OSs is lacking.

In this work, we propose a spectroscopic approach for quantification of the dynamic disorder and assessment of $\mu$. We introduce spectroscopic mobility, $\mu_s$, and show that it clearly correlates with device $\mu$ for high- and low-mobility OSs; the temperature dependencies of $\mu_s$ and $\mu$ for band-like OSs correlate as well. The obtained results highlight the efficiency of the suggested approach for screening high-$\mu$ OSs among a plenty of available materials.

## 2. Spectroscopic parameter of order in OS

The suggested approach is based on a general idea that the LF Raman spectrum of an OS crystal contains the information about dynamic disorder in it as illustrated in **Figure 1.** The Raman intensity, $I$, is determined by the vibrationally induced fluctuations (modulation) of the material polarizability, $\alpha$:[37-41]

$$I \sim \left\langle \left| \delta\alpha^2 \right| \right\rangle \equiv \left\langle \left| \left( \alpha - \alpha_0 \right)^2 \right| \right\rangle, \tag{1}$$

where $<\ldots>$ denotes ensemble averaging, and $\alpha_0$ is the polarizability in the absence of the fluctuations. According to the quantum-mechanical Kramers-Heisenberg-Dirac equation,[42-44] $\alpha$ is related to the energies and dipole moments of the transitions between the ground and various excited electronic states. Expanding $\alpha$ in Taylor series over dimensionless vibrational displacements, $q_i$, and considering the (pre-)resonance case, where the contribution of the transition from the ground state to the lowest dipole-allowed excited state dominates, we obtain the following expression for the Raman intensity of vibrational mode $i$:

$$I_i \sim \left(\frac{\partial E_1}{\partial q_i}\right)^2 \left\langle q_i^2 \right\rangle \frac{\left(\left(\mathbf{e}_L \mathbf{D}_{01}\right)\left(\mathbf{e}_S \mathbf{D}_{10}\right)\right)^2}{\left(\left(E_1 - E_0 - \hbar\omega_L - \hbar\omega_i\right)^2 + \Gamma^2\right)^2} = X \cdot Y,$$ (2)

$$X = \left(\frac{\partial E_1}{\partial q_i}\right)^2 \left\langle q_i^2 \right\rangle, \quad Y = \frac{\left(\left(\mathbf{e}_L \mathbf{D}_{01}\right)\left(\mathbf{e}_S \mathbf{D}_{10}\right)\right)^2}{\left(\left(E_1 - E_0 - \hbar\omega_L - \hbar\omega_i\right)^2 + \Gamma^2\right)^2},$$

where $\omega_i$ is the frequency of the mode, $\omega_L$ is the pump (incident light) frequency, $E_0$ and $E_1$ are respectively the energies of the ground and the lowest dipole-allowed excited state, $\mathbf{D}_{01}$ and $\mathbf{D}_{10}$ are correspondingly the dipole moments of the transitions from the ground to the lowest excited state and vice versa, $\hbar$ is the reduced Planck constant, $\mathbf{e}_L$ and $\mathbf{e}_S$ are respectively the polarizations of the incident and scattered light, and $\Gamma$ is the excited state bandwidth. The transition dipole moments and state energies are taken at the equilibrium geometry ($q_i$=0). The details of Eq. (2) derivation and used approximations are given in Supporting Information (SI), Section 1.1.

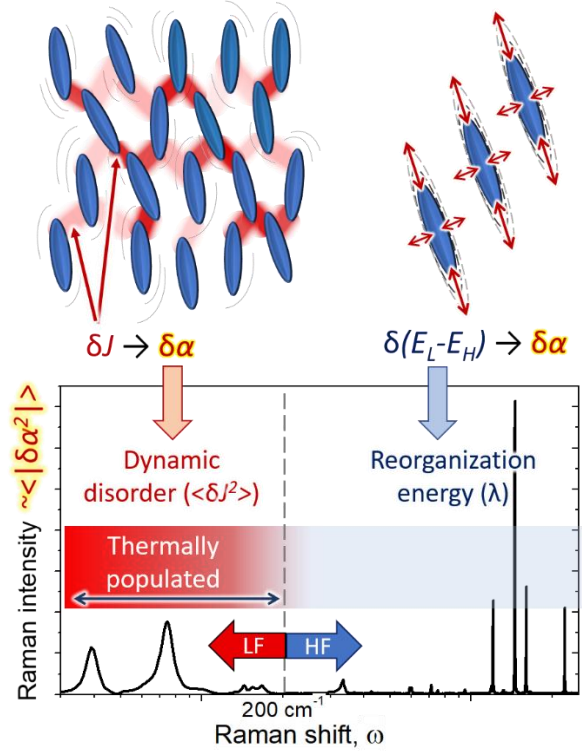

Figure 1. Illustration of the relationship between the dynamic disorder and LF Raman spectrum. The higher the intensity of LF Raman as compared to the HF one, the stronger the dynamic disorder.

Eq. (2) states that the Raman signal from $i$-th vibrational mode is determined by the two factors: $X$ that corresponds to electron-phonon interaction, and $Y$ that corresponds to light-matter interaction. The $X$-factor is related to charge transport and is in the focus of the current study. In one-dimensional OS crystal, the energy of the lowest excited state required for Eq. (2) can be approximated as $E_1 \approx \varepsilon - 2J_h - 2J_e$, where $\varepsilon$ is the difference between the energies of the highest occupied molecular orbital (HOMO) and the lowest unoccupied molecular orbital (LUMO), while $J_e$ and $J_h$ are electron and hole transfer integrals, respectively.[4] LF vibrations modulate mostly $J$,[7] and hence the LF Raman intensity is related to $\sigma_J$, i.e., dynamic disorder:

$$I \sim \sum_i \left( \frac{\partial J}{\partial q_i} \right)^2 \left\langle q_i^2 \right\rangle \cdot Y = \sigma_J^2 \cdot Y \,, \tag{3}$$

where summation runs over all LF vibrational modes, and the dependence of $Y$ on $\omega_i$ is neglected. For simplicity, the vibrational modulations of $J_e$ and $J_h$ are considered equal: $\frac{\partial J}{\partial q} = \frac{\partial J_h}{\partial q} + \frac{\partial J_e}{\partial q} = 2 \frac{\partial J_h}{\partial q}$. The HF vibrations modulate mostly $\varepsilon$;[7] therefore, the HF Raman intensity is related to the local electron-phonon interaction:[35]

$$I \sim \sum_i \left( \frac{\partial \varepsilon}{\partial q_i} \right)^2 \left\langle q_i^2 \right\rangle Y = \sum_i 4 \omega_i \lambda_i \cdot Y \,, \tag{4}$$

where $\lambda_i = \frac{1}{2\omega_i} \left( \frac{\partial \varepsilon}{\partial q_i} \right)^2$ is the contribution of the $i$-th vibrational mode to the reorganization energy.[4] Details of Eqs. (3,4) derivations and three-dimensional case are given in SI, Section 1.2. In Eq. (4), $\left\langle q_i^2 \right\rangle = 1$ since the HF vibrations are not thermally populated, and the variance of $q_i$ is determined by zero-point oscillations, making the HF Raman intensity nearly temperature-independent.[35] In contrast, the LF vibrations are populated at room temperature, and we expect decrease of LF Raman intensity with cooling due to the decrease of $\left\langle q_i^2 \right\rangle$.[35] To summarize, Eqs. (3) and (4) show that the LF Raman intensity is related to the dynamic disorder (non-local electron-phonon interaction), while the HF Raman intensity is associated with the local electron-phonon interaction. This allows us to introduce a dimensionless spectroscopic criterion, $\xi$, as an experimental parameter of order in OS:

$$\xi = \frac{\int\limits_{HF} (I/\omega) d\omega}{\int\limits_{LF} (I/\omega) d\omega} \,. \tag{5}$$

The Raman spectrum divided by $\omega$ will be referred hereafter as reduced Raman spectrum, while $\xi$ will be referred as spectroscopic order parameter. From Eqs. (3–5), it follows that

$$\xi \sim \frac{\lambda \cdot Y}{\sum_{LF} \omega_i^{-1} \left(\partial J / \partial q_i\right)^2 \left\langle q_i^2 \right\rangle \cdot Y} = \frac{\lambda}{\sigma_J^2 \tau_{eff}}, \qquad (6)$$

where $\tau_{eff}$ is the average period of LF vibrations weighted over their contribution to $J$ modulation. For given $\lambda$ (which can be readily calculated using DFT[1,2]), the weaker the dynamic disorder and the shorter the $\tau_{\text{eff}}$, the higher the $\xi$ value. Importantly, in contrast to the Raman intensities, the $\xi$ value is nearly insensitive to experimental conditions such as the sample thickness, pump intensity, detector sensitivity, as well as parameters of light-matter interaction such as the transition dipole moment and the proximity of the pump photon energy to the resonance ($Y$ in Eq. 4; see SI, Section 1.3). Thus, we conclude that $\xi$ can be used to quantify and compare dynamic disorder in various OSs.

### 3. Spectroscopic assessment of charge transport

The key claim of this study is that the suggested spectroscopic order parameter ($\xi$) allows estimation of $\mu$. According to the transient localization scenario of charge transport,[8,9] $\mu$ is determined by the (transient) delocalization length of a charge carrier, $L_{\text{D}}$, and the period of the characteristic LF vibration, $\tau$:

$$\mu = \frac{e}{kT} \frac{L_D^2}{2\tau}, \qquad (7)$$

where $e$ is the elementary charge, $k$ is the Boltzmann constant, $T$ is the absolute temperature, and the one-dimensional case is addressed. It is reasonable to assume that $L_D$ increases with $J$ and decreases with the dynamic disorder quantified by $\sigma_J$. According to Ref. [45], $L_D \sim a \left(J / \sigma_J\right)^x$, where $a$ is the distance between the sites (molecules), and $x$ varies in the range 2/3–4/3. In Ref.

[8], fitting the experimental data by $L_D \sim a\left(J^m / \sigma_J^n\right)$ yielded $m\sim1$ and $n\sim1.6$. Considering these uncertainties in $x$, we assume $L_D = a\left(J / \sigma_J\right)$ for simplicity; and presuming $\tau \approx \tau_{eff}$ obtain

$$\mu = \frac{e}{2kT} \frac{J^2 a^2}{\sigma_J^2 \tau} \qquad (8),$$

According to Eq. (8), $\mu$ strongly depends on $\sigma_J$ and $\tau$. Heavy impact of the dynamic disorder on $\mu$ also follows from the band model, which is widely used for description of charge transport in high-$\mu$ OSs (See SI, Section 1.4). Using the concept of spectroscopic order parameter and following Eqs. (6, 8), we obtain:

$$\mu \approx \frac{e}{2kT}\left(\frac{a^2 J^2}{\lambda}\right)\xi \qquad , \qquad (9)$$

i.e., $\mu$ is proportional to $\xi$. We thus introduce the "spectroscopic mobility" for quantitative evaluation of charge transport:

$$\mu_S = C\,\frac{e}{kT}\left(\frac{\sum_j a_j^2 J_j^2}{\lambda}\right)\frac{\int\limits_{HF}\left(I/\omega\right)d\omega}{\int\limits_{LF}\left(I/\omega\right)d\omega}, \qquad (10)$$

where $C$ is a dimensionless constant, and summation runs over all charge transport directions. The $\mu_s$ requires the Raman spectrum, either experimental or theoretical (e.g., calculated using solid-state DFT),[46,47] as well as $J$ and $\lambda$ values, which can be calculated using DFT.[1] In high-$\mu$ OSs with coherent charge transport, for which Eq. (7) applies,[8,10] $\mu_s$ is expected to correlate with $\mu$ and can be used for its rapid estimation. Moreover, since $\mu$ in the hopping regime is also sensitive to $J$, $\lambda$ and $\sigma_J$,[48] we expect that $\mu_s$ correlates with $\mu$ for low-$\mu$ OSs as well. Although $\mu_s$ was introduced under rather strong approximations, its efficiency is *a posteriori* justified in the following section by the experimental data.

### 4. Experimental verification of the approach

#### 4.1. Temperature dependence

First of all, we investigated temperature dependencies of Raman spectra for three OSs, in which $\mu$ increases with cooling: 2,5-difluoro-7,7,8,8-tetracyanoquinodimethane (F$_2$-TCNQ),[49] rubrene,[50] and naphthalene.[51] Increase of $\mu$ with cooling is commonly assigned to coherent (band-like) charge transport,[1,2,4] and we expected correlation between the temperature dependencies of $\xi$ derived from the Raman spectra and $\mu$. As the lower bound of the HF range, we used 600 cm$^{-1}$. **Figure 2a-c** shows recorded LF Raman spectra for the studied materials at various temperatures. Cooling results in significant increase in the frequencies of LF modes and dramatic decrease in their integral intensities. The former is commonly attributed to crystal shrinking, which should strengthen intermolecular interactions.[25,52] In contrast, the dramatic decrease in Raman *intensities* of LF vibrations with cooling for OSs was rarely noticed earlier.[53] For F$_2$-TCNQ (Figure 2a), the lowest-frequency mode undergoes the most pronounced suppression: its integral intensity drops in 5.9 times from 398 K down to 83 K, while the integral intensities of the other two modes decrease in 3.8 and 2.4 times. Similar to F$_2$-TCNQ, the lowest-frequency mode in rubrene and naphthalene exhibits the strongest intensity decrease with cooling (Figure 2b,c). The HF Raman intensities show negligible changes with temperature.

The abovementioned increase of $\omega_i$ and decrease of $I_i$ for the LF modes (Figure 2a-c) along with negligible changes of the HF Raman spectrum promote a pronounced increase in $\xi$ with cooling as shown in Figure 2d. According to the suggested approach, this increase indicates reduction of dynamic disorder and hence weakening of the impact of LF vibrations on charge transport. The latter is quite natural and is in accordance with theoretical results from Refs. [54,55] and experimental findings of Ref. [53]. The increase of $\mu$ with cooling in OSs with band-like charge transport can be explained by suppression of $\sigma_J$ as follows from Eqs. (8). Figure 2d shows that $\xi$ clearly correlates with the OFET hole mobility for rubrene and electron mobility for $F_2$-TCNQ down to 150 K. The lack of correlation between the OFET $\mu$ and $\xi$ below 150 K can be attributed to domination of incoherent charge transport due to increased role of charge traps; therefore, the OFET $\mu$ no more represents the intrinsic charge mobility in this temperature range.[56,57] Indeed, the Hall mobility, which is sensitive only to the coherent transport, was shown to further increase with cooling in rubrene crystals even when the OFET $\mu$ started to decrease.[58] The charge mobility in ultrapure naphthalene single crystals measured using the time-of-flight (ToF) technique[51] perfectly correlates with $\xi$ as well. The correlation between the temperature dependencies of $\mu$ and $\mu_s$ is presented in SI, Figure S2.

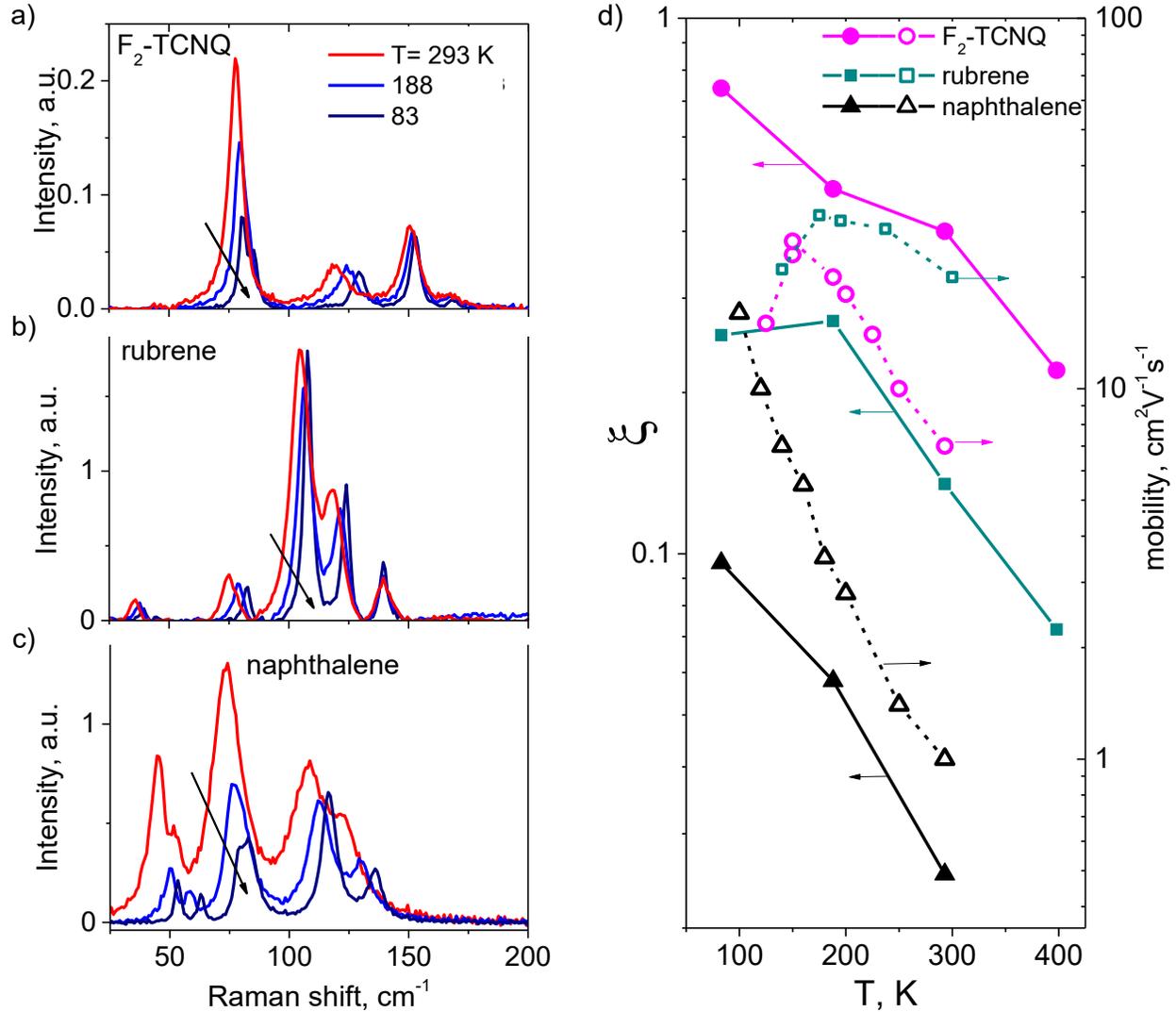

**Figure 2.** Effect of temperature on LF Raman spectra (a-c), charge mobility, $\mu$, and spectroscopic order parameter, $\xi$, (d) in F$_2$-TCNQ, rubrene, and naphthalene. Experimental Raman spectra of F$_2$-TCNQ (a, polycrystalline powder), rubrene (b, single crystal), and naphthalene (c, polycrystalline powder) are normalized to the maximal intensity in the HF range. The arrows highlight decrease of the integral Raman intensity with cooling. The OFET $\mu$ data for single crystals of F$_2$-TCNQ and rubrene in panel (d) are from Refs. [49] and [50], and the time-of-flight $\mu$ data for naphthalene single crystals are from Ref. [51].

### 4.2. Various materials

To test the efficiency of $\xi$ and $\mu_s$ for evaluation of the dynamic disorder and intrinsic charge mobility for different materials, we applied our approach to a set of popular OSs. As shown in Ref. [59], the pure hopping mobility does not exceed 0.1 cm$^2$V$^{-1}$s$^{-1}$, and it is commonly believed that much higher $\mu$ is a signature of significant contribution from coherent charge transport. We therefore chose the materials for which $\mu \geq 0.5$ cm$^2$V$^{-1}$s$^{-1}$ was reported: rubrene, F$_2$-TCNQ, 7,7,8,8-tetracyanoquinodimethane (TCNQ), tetracene, anthracene, naphthalene, and tetramethyl-tetrathiafulvalene (TM-TTF). For comparison, we also studied several relatively low-mobility materials ($\mu < 0.5$ cm$^2$/Vs): 2,2':5',2'':5'',2'''-quarterthiophene (4T), 2,3,5,6-tetrafluoro-7,7,8,8-tetracyanoquinodimethane (F$_4$-TCNQ), 5,5'-diphenyl-2,2'-bithiopene (PTTP), 1,4-bis(5-phenyl-2-thienyl)benzene) (AC5), p-terphenyl (3P), p-tetraphenyl (4P), and bis(ethylenedithio)tetrathiafulvalene (BEDT-TTF). As an approximation to the OS intrinsic mobility, we used the OFET values if the data were available and the ToF ones otherwise. For the OFET $\mu$, the highest value reported on either single crystal or thin film samples was used, but the data showing the measurement reliability factor $r < 50\%$[3] were not taken into account. The $\mu$ data in the linear regime of OFET, if available, were used as the most reliable approximation to the intrinsic mobility.[3] The details of the used $\mu$ data are given in Table S1. To record high-quality LF Raman spectra of OSs, we studied only the compounds that do not strongly absorb and luminesce at the Raman excitation wavelength used (633 nm).

**Figure 3a** compares reduced LF Raman spectra of the investigated high-$\mu$ OSs. The as recorded Raman spectra are given in Figure S2. OSs with the higher $\mu$ (e.g., rubrene and F$_2$-TCNQ) typically show the lower reduced Raman intensity in the LF range than those with lower $\mu$ (e.g., naphthalene and TM-TTF), in line with our assumption that an intensive LF Raman spectrum implies strong dynamic disorder. **Figure 3b** illustrates the correlation between $\xi$ obtained from the Raman data and an estimate of the dynamic disorder calculated using molecular dynamics in Ref. [48] — paracrystalline (dis)order parameter $g = \Delta/d_0$, where $d_0$ stands for the average distance between the centers of the molecules, and $\Delta = \sqrt{\langle d^2 \rangle}$ is the standard deviation for this distance. Figure 3b shows that the higher the $1/g$ value (weaker dynamic disorder), the higher $\xi$. This correlation corroborates our claim that high $\xi$ reveals OSs with small thermally-induced molecular displacements and hence weak dynamic disorder.

**Figure 4a** shows correlation between the $\xi$ and room-temperature $\mu$ values for various OSs. Remarkably, for high-$\mu$ OSs ($\mu$>0.5 cm$^2$/Vs, points on the right of vertical red line in Figure 3a), $\mu$ is nearly proportional to $\xi$. The clear correlation between $\xi$ and $\mu$ for these OSs supports the efficiency of our spectroscopic order parameter and corroborates the suggestion that dynamic disorder limits coherent charge transport in high-$\mu$ OSs.[8,10] Lack of correlation between $\mu$ and $\xi$ in the whole investigated set of OSs (i.e., including low-$\mu$ materials – points on the left of the vertical red line in Figure 4a) can be attributed to the impact of $J$ and $\lambda$ on $\mu$ as predicted by Eq. (9). Indeed, both $J$ and $\lambda$ are comparable for various high-$\mu$ OSs,[2] making $\xi$ the factor limiting charge transport. However, for low-$\mu$ OSs, low $J$ and/or large $\lambda$ prevent coherent charge transport and become the main hindrances for high $\mu$. In contrast to $\xi$, $\mu_S$, which takes into account $J$ and $\lambda$ according to Eq. (10), nicely correlates with $\mu$ for the whole set of OSs including high- and low-$\mu$ ones as presented in Figure 4b. Some deviations (e.g. for 4P and 3P) could be attributed to non-optimized devices with underestimated $\mu$ values.

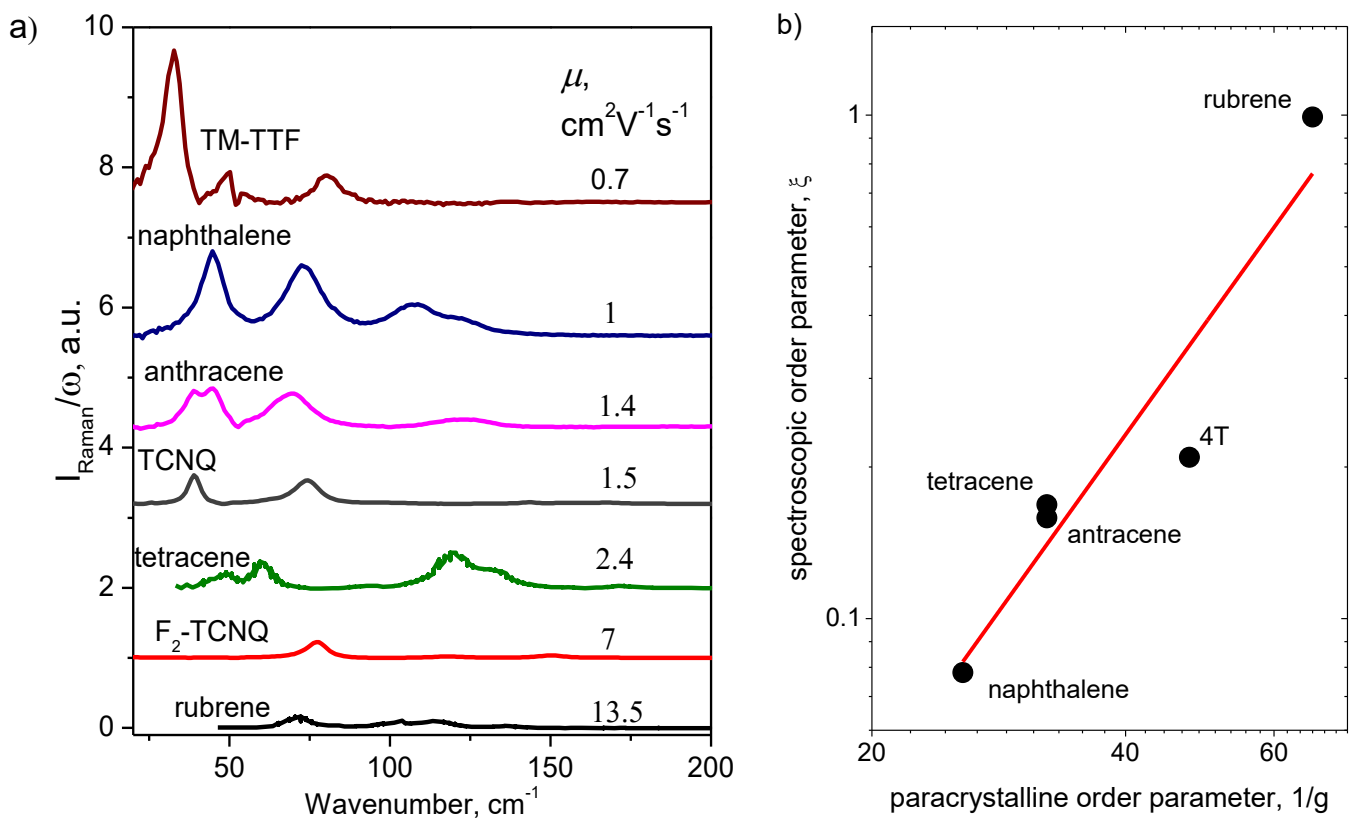

**Figure 3**. Low-frequency reduced Raman spectra (a) and correlation between the spectroscopic order parameter, ξ, and the calculated paracrystalline order parameter, $1/g$ (b). Spectra at panel (a) are arranged according to the maximal reported $\mu$ from top (lowest) to bottom (highest). As-recorded Raman spectra (SI, Figure S1) were normalized to the maximum in the HF part and then divided by the wavenumber. All the spectra except that of rubrene and tetracene were measured in polycrystalline powders to avoid the effect of anisotropy and polarization. The rubrene and tetracene spectra were taken from Ref. [22]. The $g$ values in panel (b) are taken from Ref. [48], and the line is a linear fit.

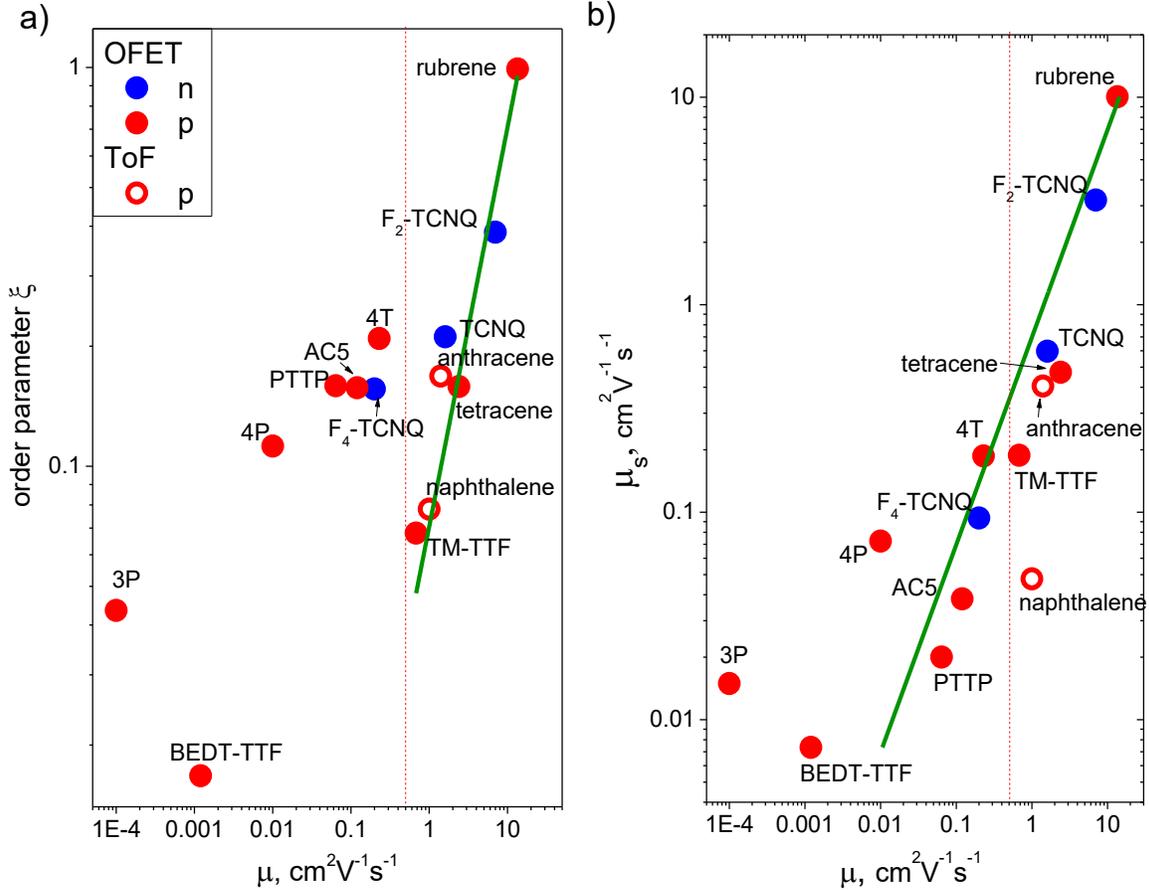

**Figure 4**. Correlation between the spectroscopic order parameter, $\xi$, and charge mobility, $\mu$ (a), and correlation between spectroscopic mobility $\mu_s$ and $\mu$ (b). The red line separates the high-$\mu$ OSs (right) from the low-$\mu$ ones (left). The green lines are linear fits with zero intercept for high-$\mu$ OSs (a) and all OSs (b). The difference in the room-temperature $\xi$ for rubrene in Figures 2 and 4 results from the different experimental conditions (one orientation of the single crystal in Figure 2 and averaged over multiple orientations of single crystals in Figure 4). The $\mu$ data were taken from Refs. [49-51, 60-68]. Filled circles correspond to OFET $\mu$ values, open ones correspond to ToF $\mu$ values. For details, see SI, Section 4.

## 5. Discussion

The established correlations (Figures 2d, 3b and 4) confirm that the introduced spectroscopic order parameter, $\xi$, and spectroscopic mobility, $\mu_s$, are efficient tools for rapid estimation of the dynamic disorder and intrinsic charge mobility, respectively. Based on our results, we propose the following protocol for screening the high-mobility OSs. This protocol is visualized in **Figure 5.** First, the Raman spectrum is recorded, and $\xi$ is calculated according to Eq.

(5). If the $\xi$ value is high (e.g., exceed 0.1), this material is relatively rigid, and is worth further investigation, i.e., calculation of $\mu_s$ using Eq. (10). For this purpose, $\lambda$ and $J$ values are necessary. While the former can be readily obtained from single-molecule DFT calculations,[1,2] $J$ estimation requires the crystal structure, which usually needs X-ray analysis. Nevertheless, Raman and X-ray experiments followed by $\lambda$ and $J$ calculations are considerably more robust, faster and simpler than device optimization for reliable $\mu$ measurements. Therefore, $\mu_s$ is a promising tool for revealing candidates for high-$\mu$ OSs. Successive measurements of $\mu$ in devices (e.g., in OFETs) would judge which of these candidates indeed support efficient charge transport.

Apart the suggested practical application, the presented results are also of fundamental importance since they establish the relationship between optical properties (Raman spectrum), structural properties (dynamic disorder) and charge transport properties ($\mu$). This relationship enables studies of the impact of various vibrational modes on charge transport and its anisotropy[6] under various conditions, e.g., temperature (see Figure 1), pressure and mechanical strain. The results of such studies can provide a deep insight into the factors governing charge transport in OSs, which is necessary for refinement of charge transfer models.[8-11]

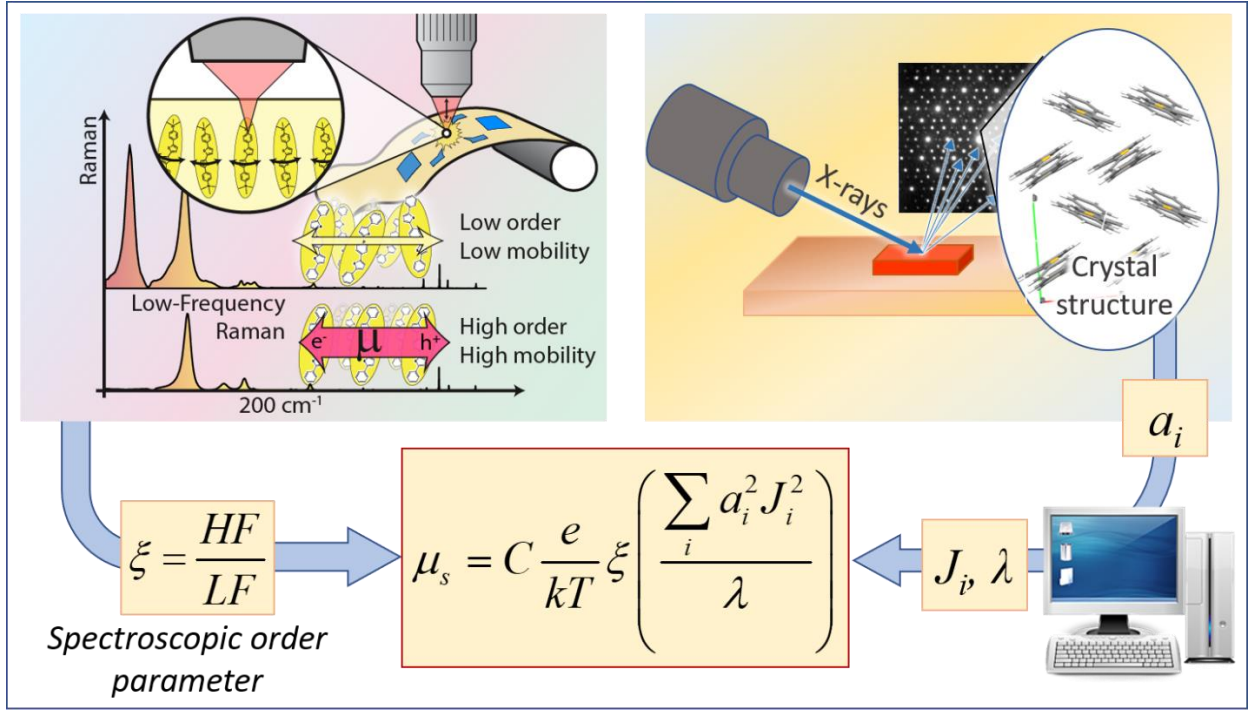

**Figure 5.** The proposed protocol for screening of high-$\mu$ OSs.

## 6. Conclusions

To summarize, we have suggested the spectroscopic approach for evaluation of charge mobilities in OSs prior their measurements in electronic devices. The approach is based on our finding that, in OSs, the low-frequency (<200 cm$^{-1}$) Raman intensity is linked to the dynamic disorder. The spectroscopic mobilities clearly correlate with the device ones in various OSs. The temperature dependencies of the spectroscopic mobility and band-like device mobility correlate as well. Being rapid and robust, the spectroscopic assessment of charge mobility allows screening of high-mobility OSs among the plethora of available materials. We thus anticipate that the suggested approach provides efficient guidelines for the focused search of promising OSs and will boost the development of new high-mobility OSs.

## 7. Experimental and Computational

Powders of TCNQ (Sigma-Aldrich), $F_2$-TCNQ (TCI company), naphthalene (Merck) and anthracene (Merck) were used as received. TM-TTF and BEDT-TTF were synthesized as described in Refs. [69] and [70], respectively. 4T and PTTP were synthesized according to the procedures described in Refs. [71] and [72], correspondingly.

Raman measurements for TCNQ, $F_2$-TCNQ, naphthalene, anthracene and TM-TTF powders and rubrene single crystals were conducted using a Raman microscope (inVia, Renishaw) with an excitation wavelength of 633 nm provided by a He-Ne laser. LF measurements were conducted using NExT monochromator, HF measurements were performed with a 633-nm edge filter. The LF and HF measurements were performed separately, and then the LF and HF spectra were merged. All spectra for powder samples were measured in several points and then averaged to reduce the anisotropy effect on the Raman spectra. Raman spectra at various temperatures were recorded with the use of temperature-controlled microscope stage (THMS600, Linkam).

The $J$ values required for $\mu_s$ were calculated according to the procedure described in Ref. [73], and $\lambda$ values were obtained according to the 4-point scheme.[1] All DFT calculations were performed using GAMESS package[74,75] at B3LYP/6-31g(d) level.

## Electronic supplementary material

Electronic supplementary material is available.

## Conflicts of interest

There are no conflicts to declare.

**Acknowledgements**

The work on high-mobility OSs was supported by Russian Foundation for Basic Research (project #16-32-60204 mol_a_dk). The work on low-mobility OSs and calculation of $J$ and $\lambda$ were supported by Russian Science Foundation (project #18-72-10165). The study was conducted using equipment purchased under the Lomonosov Moscow State University Program of Development. The authors thank V. Postnikov and E. Komissarova for their assistance in preparation of the OS samples and N. Melnik for his advice on Raman measurements in rubrene single crystals.

**Supporting Information**

## 1. Model formulation: details

### 1.1. Raman intensities

In this section, we provide a detailed derivation of Eq. (5) that links the frequencies, Raman intensities, and contributions to the electron-phonon coupling for various vibrational modes. According to the quantum-mechanical approach, the Raman scattering intensity is[1]

$$I \sim \left| \left\langle \varphi_0^{fin} \left| \hat{\alpha}_{\alpha\beta} \right| \varphi_0^{in} \right\rangle \right|^2 \tag{S1}$$

where $\left| \varphi_0^{in} \right\rangle$ and $\left| \varphi_0^{fin} \right\rangle$ are the wavefunctions of the initial and final vibrational states of the electronic ground state, and $\hat{\alpha}_{\alpha\beta}$ is the polarizability tensor. The $\hat{\alpha}_{\alpha\beta}$ is related to the energies $E_k$ and dipole moments $\boldsymbol{D}_{0k}$ of the transitions between the ground state and $k$-th excited electronic state[1] according to the Kramers-Heisenberg-Dirac equation[2]:

$$\alpha_{LS} = \sum_k \sum_i \left( \frac{(\mathbf{e}_L \mathbf{D}_{0k}) \left| \varphi_k^i \right\rangle \left\langle \varphi_k^i \right| (\mathbf{e}_S \mathbf{D}_{k0})}{E_k - E_0 - \hbar\omega_i - \hbar\omega_L - i\Gamma} + \frac{(\mathbf{e}_S \mathbf{D}_{0k}) \left| \varphi_k^i \right\rangle \left\langle \varphi_k^i \right| (\mathbf{e}_L \mathbf{D}_{k0})}{E_k - E_0 - \hbar\omega_i + \hbar\omega_S + i\Gamma} \right) \tag{S2}$$

where $\omega_i$ is the frequency of the mode, $\omega_L$ is the pump frequency, $E_0$ is the energy of the ground state, $\hbar$ is the reduced Planck constant, $\mathbf{e}_L$ and $\mathbf{e}_S$ are respectively the polarizations of the incident (laser, $L$) and scattered ($S$) light, $\Gamma$ is the excited state bandwidth, $\left| \varphi_k^i \right\rangle$ is the $i$-th vibrational function of the $k$-th electronic state. The transition dipole moments and state energies are taken at the equilibrium geometry ($q_i$=0).

Consider the pre-resonance regime of Raman scattering, i.e., when the incident photon energy is close to the energy of the lowest dipole-allowed excited electronic state, $k$=1. In this case, the contribution of the latter state to the Raman signal is dominant, and the first term in Eq. (S2) is considerably larger than the second one. Thus,

$$\alpha_{LS} = \sum_i \frac{(\mathbf{e}_L \mathbf{D}_{01})(\mathbf{D}_{10} \mathbf{e}_S)}{E_1 - E_0 - \hbar\omega_i - \hbar\omega_L - i\Gamma} \tag{S3}$$

Below, we will address the isotropic Raman signal and denote $D_{01} = (\mathbf{e}_L \mathbf{D}_{01})$, $D_{10} = (\mathbf{e}_S \mathbf{D}_{10})$.

Expanding $\alpha$ in Taylor series over dimensionless coordinates $q_i$, $\alpha = \alpha_0 + \sum_i q_i \left. \frac{\partial \alpha}{\partial q_i} \right|_{q=0}$, we obtain

$$\alpha - \alpha_0 \approx \sum_i -\frac{\left(D_{10}D_{01}\right)\left(\dfrac{\partial\left(E_1-E_0\right)}{\partial q_i}\right)}{\left(E_1-E_0-\hbar\omega_i-\hbar\omega_L-i\Gamma\right)^2}q_i + \frac{\left(D_{10}\right)\left(\dfrac{\partial D_{01}}{\partial q_i}\right)+\left(D_{01}\right)\left(\dfrac{\partial D_{10}}{\partial q_i}\right)}{E_1-E_0-\hbar\omega_i-\hbar\omega_L-i\Gamma}q_i \qquad (S4)$$

In Eq. (S4), $D_{01}$, $D_{10}$ and $E_1$, $E_0$ are taken at equillium ($q_i$=0). For OSs in the pre-resonance regime, the first term in the right hand side of Eq. S4 (corresponding to Albrecht A-term, or Franck-Condon term) is considerably larger than the second one (corresponding to Albrecht B-term, or Herzberg-Teller term).[1] In these approximations, the Raman intensity of the $i$-th mode reads:

$$I_i \propto \left(\frac{\partial E}{\partial q_i}\right)^2 \left\langle q_i^2\right\rangle \cdot \frac{\left(D_{10}D_{01}\right)^2}{\left(\left(E-\hbar\omega_i-\hbar\omega_L\right)^2+\Gamma^2\right)^2} = \left(\frac{\partial E}{\partial q_i}\right)^2\left\langle q_i^2\right\rangle \cdot Y, \qquad (S5)$$

where $E=E_1-E_0$ and $Y = \dfrac{\left(D_{10}D_{01}\right)^2}{\left(\left(E-\hbar\omega_i-\hbar\omega_L\right)^2+\Gamma^2\right)^2}$ describes light-matter interaction.

Noteworthily, Eq. (S5) for Raman intensity nearly coincides with that obtained from the widely used excited-state-gradient approximation[3].

Eq. (S5) is exactly Eq. (2) of the main text.

### 1.2. Modulation of the excited state energy by low- and high-frequency vibrations

As follows from Eq. (S5), the Raman intensity of a given mode is governed by its ability to modulate the energy of the lowest excited dipole-allowed state. In an isolated molecule, the latter can be roughly described as an electron at the LUMO and a hole at the HOMO with the energy[4]

$$E = E_L - E_H - E_{b1} = \varepsilon - E_{b1} \qquad (S6)$$

where $E_L$ is the LUMO energy, $E_H$ is the HOMO energy, $E_{b1}$ is the energy of the electron-hole interaction, and $\varepsilon = E_L - E_H$ is the HOMO-LUMO gap. Accordingly, the impact of vibrations on $E$ can be divided into the modulations of $\varepsilon$ and $E_{b1}$:

$$\frac{\partial E}{\partial q} = \frac{\partial \varepsilon}{\partial q} - \frac{\partial E_{b1}}{\partial q}. \qquad (S7)$$

Since $E_{b1}$~0.5 eV$<<\varepsilon$~3 eV for typical OSs, we assume that $E_{b1}$ — the Coulomb binding energy — weakly depends on intramolecular vibrations, resulting in $\dfrac{\partial \varepsilon}{\partial q} >> \dfrac{\partial E_{b1}}{\partial q}$, and hence $\dfrac{\partial E}{\partial q} \approx \dfrac{\partial \varepsilon}{\partial q} = \dfrac{\partial E_L}{\partial q} - \dfrac{\partial E_H}{\partial q}$. Further, we assume $\dfrac{\partial E_L}{\partial q} = -\dfrac{\partial E_H}{\partial q} = \dfrac{1}{2}\dfrac{\partial \varepsilon}{\partial q}$: vibrations that modulate intramolecular conjugation affect both $E_L$ and $E_H$, i.e., alter $\varepsilon$. This is in line with the results of Ref. [5].

In crystal, HOMOs of the molecules interact and form the valence band. Similarly, LUMOs form the conduction band. In the first excited state, the hole resides at the top of the valence band, and its energy can be approximated by $E_v = E_H + 2J_h$, where $J_h$ is the hole transfer integral. The electron resides at the bottom of the conduction band ($E_C$), and its energy is $E_C = E_L - 2J_e$, where $J_e$ is the electron transfer integral. The energy of the first excited state of an infinite one-dimensional OS crystal, i.e. the exciton, is

$$E \approx \varepsilon - 2J_{exc}, \tag{S8}$$

where $J$ is the exciton transfer integral. Since the exciton in OSs is commonly assumed localized at one molecule (Frenkel-type)[6], we consider $E_b \sim E_{b1}$ and neglect $\dfrac{\partial E_b}{\partial q}$ for crystal as was done for isolated molecule (see above). We further assume that LF vibrations affect $J_h$ and $J_e$ in similar way: $\dfrac{\partial J_h}{\partial q} \sim \dfrac{\partial J_e}{\partial q}$. Therefore, $\dfrac{\partial E}{\partial q} \approx \dfrac{\partial \varepsilon}{\partial q} - 2\left(\dfrac{\partial J_h}{\partial q} + \dfrac{\partial J_e}{\partial q}\right) \approx 2\left(\dfrac{\partial E_H}{\partial q} - 2\dfrac{\partial J_h}{\partial q}\right)$. Within the mentioned approximations, the Raman intensity for a given mode of OS crystal reads:

$$I_i \sim \left(2\dfrac{\partial E_H}{\partial q_i} - 4\dfrac{\partial J_h}{\partial q_i}\right)^2 \langle q_i^2\rangle \cdot Y = 4\cdot\left[\left(\dfrac{\partial E_H}{\partial q_i}\right)^2 - 4\left(\dfrac{\partial E_H}{\partial q_i}\right)\left(\dfrac{\partial J_h}{\partial q_i}\right) + 4\left(\dfrac{\partial J_h}{\partial q_i}\right)^2\right]\langle q_i^2\rangle \cdot Y, \tag{S9}$$

For three-dimensional crystal, Eq. (S8) transforms to $E = \varepsilon - 2\sum_{j=1}^{N/2}\left(J_e^{\ j} + J_h^{\ j}\right) - E_b$, where $N$ is the number of molecules adjacent to a given one, and summation runs over all the nonequivalent electron (hole) transfer integrals $J_e^j$ ($J_h^j$) between a given molecule and its neighbors. Eq. (S9) in this case transforms to:

$$I_i \sim \left[\left(\dfrac{\partial E_H}{\partial q_i}\right)^2 - 4\left(\dfrac{\partial E_H}{\partial q_i}\right)\sum_{j=1}^{N/2}\left(\dfrac{\partial J_h^{\ j}}{\partial q_i}\right) + 4\sum_{l=1}^{N/2}\sum_{j=1,j\neq l}^{N/2}\left(\dfrac{\partial J_h^{\ j}}{\partial q_i}\right)\left(\dfrac{\partial J_h^{\ l}}{\partial q_i}\right) + 4\sum_{j=1}^{N/2}\left(\dfrac{\partial J_h^{\ j}}{\partial q_i}\right)^2\right]\langle q_i^2\rangle \cdot Y$$
$$\tag{S10}$$

Large $\left(\dfrac{\partial E_H}{\partial q_i}\right)$ and $\left(\dfrac{\partial J_h}{\partial q_i}\right)$ are usually observed for HF and LF vibrations, correspondingly.[7] The term $\left(\dfrac{\partial E_H}{\partial q_i}\dfrac{\partial J_h}{\partial q_i}\right)$ is hence expected to be low due to either low $\left(\dfrac{\partial E_H}{\partial q_i}\right)$ in the LF range or low $\left(\dfrac{\partial J_h}{\partial q_i}\right)$ in the HF range,[7] and is omitted below. We also assume that the terms $\left(\dfrac{\partial J_h^{\ j}}{\partial q_i}\right)\left(\dfrac{\partial J_h^{\ l}}{\partial q_i}\right)$ are significantly lower than $\left(\dfrac{\partial E_H}{\partial q_i}\right)^2$ and $\left(\dfrac{\partial J_h}{\partial q_i}\right)^2$, and neglect the former. The above approximations allow us to simplify Eqs. (S9,10) as:

$$I_i \propto \left[ \left( \frac{\partial E_H}{\partial q_i} \right)^2 + 4 \left( \frac{\partial J_h}{\partial q_i} \right)^2 \right] q_i^2 \cdot Y . \tag{S11}$$

Eq. (S11) states that high Raman intensities are expected from Eq. (S10) for vibrational modes that strongly affect either $E_H(E_L)$ or $J_h(J_e)$.

## 1.3. Linking Raman intensities to reorganization energy and dynamic disorder

The contributions from different vibrational modes to the reorganization energy $\lambda$, which quantifies local electron-phonon interaction for electron (hole) transport, $\lambda_e$ ($\lambda_h$), are determined by $\frac{\partial E_L}{\partial Q_i}$ $\left( \frac{\partial E_H}{\partial Q_i} \right)$:[4,7]

$$\lambda_i^{e,h} = \frac{\left( \frac{\partial E_{L,H}}{\partial q_i} \right)^2}{2\omega_i} , \tag{S12}$$

In contrast, the contributions of these vibrations to lattice distortion energy $L$, which quantifies the non-local electron-phonon coupling for a three-dimensional OS crystal are determined by $\frac{\partial J_e}{\partial Q_i}$ $\left( \frac{\partial J_h}{\partial Q_i} \right)$:[4,7]

$$L_{e,h}^i = \sum_j \frac{\left( \frac{\partial J_{e,h}^j}{\partial q_i} \right)^2}{2\omega_i} . \tag{S13}$$

Combining Eqs. (S9), (S12) and (S13), we arrive to the relation that links the Raman intensity of vibrational mode $i$ to the contribution of this mode to the electron-phonon coupling:

$$\frac{I_i}{\omega_i} \propto \left( \lambda_i + 4L_i \right) \langle q_i^2 \rangle \cdot Y , \tag{S14}$$

where $\lambda_i = \lambda_e^i + \lambda_h^i$, and $L_i = L_e^i + L_h^i$. The relationship between the contributions to the electron-phonon coupling and Raman intensities in LF and HF ranges is illustrated in Fig. S1.

Since $\lambda_i$ are much larger than $L_i$ in HF range and *vice versa* in LF range,[7]

$$\xi = \frac{\left( \int_{HF} \frac{I}{\omega} d\omega \right)}{\left( \int_{LF} \frac{I}{\omega} d\omega \right)} \propto \frac{\sum_{i,HF} \lambda_i \langle q_i^2 \rangle \cdot Y}{\sum_{i,LF} L_i \langle q_i^2 \rangle \cdot Y} = \frac{\lambda}{\sum_i \frac{1}{\omega_i} \left( \frac{\partial J}{\partial q_i} \right)^2 \langle q_i^2 \rangle} = \frac{\lambda}{\sigma_J^2 \tau_{eff}} , \tag{S15}$$

where it is taken into account that $\langle q_i^2 \rangle = 1$ fo HF modes, and $\sigma_J^2 = \left\langle \left( J - J_0 \right)^2 \right\rangle$ is the dispersion of charge transfer integrals, which quantifies dynamic disorder.

Eq. (S15) coincides with Eq. (6) of the main text.

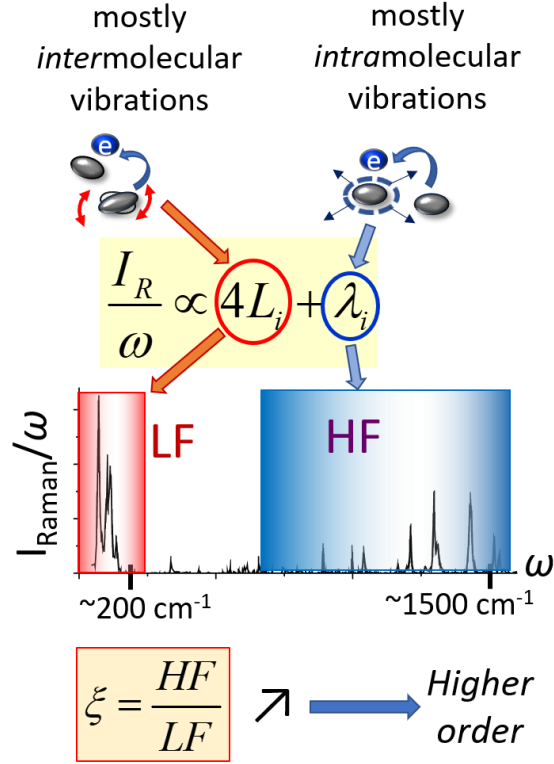

Fig. S1. Illustration of the relationship between the Raman intensity in LF and HF ranges and contributions of vibrational modes to the local ($\lambda_i$) and non-local ($L_i$) electron-phonon interaction.

### 1.4. Impact of dynamical disorder on charge mobility within band model

The impact of dynamic disorder, (quantified by $\sigma_J$), on $\mu$ within the transient localization scenario was described in Section 3 of the main text. In this section, we describe the impact of $\sigma_J$ on $\mu$ in the framework of the band model of charge transport. According to this model, $\mu$ reads:[4]

$$\mu_{band} = \frac{e\tau_s}{m^*}, \tag{S16}$$

where $\tau_S$ is the scattering time, $m^* = \dfrac{\hbar^2}{2Ja^2}$ is the effective mass of the charge carrier, and $e$ is the elementary charge. Dynamical disorder quantified by $\sigma_J$ decreases $\tau_S$:[8]

$$\tau_s \propto \frac{\hbar J}{\sigma_J^2} \tag{S17},$$

Hence, it decreases $\mu$ as well:

$$\mu_{band} = \frac{4e}{\hbar} \frac{J^2 a^2}{\sigma_J^2} \tag{S18}$$

The larger the $J$ values and the weaker their vibrational modulation (i.e., lower $\sigma_J$), the higher $\mu$. Importantly, Eq. (S18) is very similar to that obtained within the transient localization scenario, where $\mu_{TL} = \dfrac{e}{2kT} \dfrac{J^2 a^2}{\sigma_J^2 \tau}$ (see Eq. (8) of the main text). Comparison of the two expressions for charge mobility yields $\mu_{TL} = \mu_{band} \cdot \dfrac{\hbar \omega}{8kT}$, i.e., the transient localization scenario predicts stronger dependence of $\mu$ on thermal population of the lowest vibrational state than the band model, in line with Ref. [9]. Nevertheless, both models show significant dependence of $\mu$ on dynamical disorder.

### 1.5. Approximations of the proposed approach

Eqs. (2-6) of the main text are derived in the following approximations:

- pre-resonance Raman scattering regime, two-state approximation ($S_0$-$S_1$ transition has the largest contribution to the Raman signal);

- Franck-Condon approximation;

- neglect of Dushinsky mixing and Herzberg-Teller effects;

- the high-frequency (HF, $\omega > 600$ cm$^{-1}$) modes weakly contribute to the non-local coupling, while the low-frequency (LF, $\omega < 200$ cm$^{-1}$) ones weakly contribute to the local electron-phonon coupling;

- the optical bandgap is approximated as $E = \varepsilon - 2J_h - 2J_e$,

- anisotropies of Raman signal and charge mobility are neglected;

- the vibrations affect hole and electron transport equally, i.e., $\lambda_i$ and $L_i$ are the same for electrons and holes;

- excitonic effects are neglected.

**2. Raman spectra**

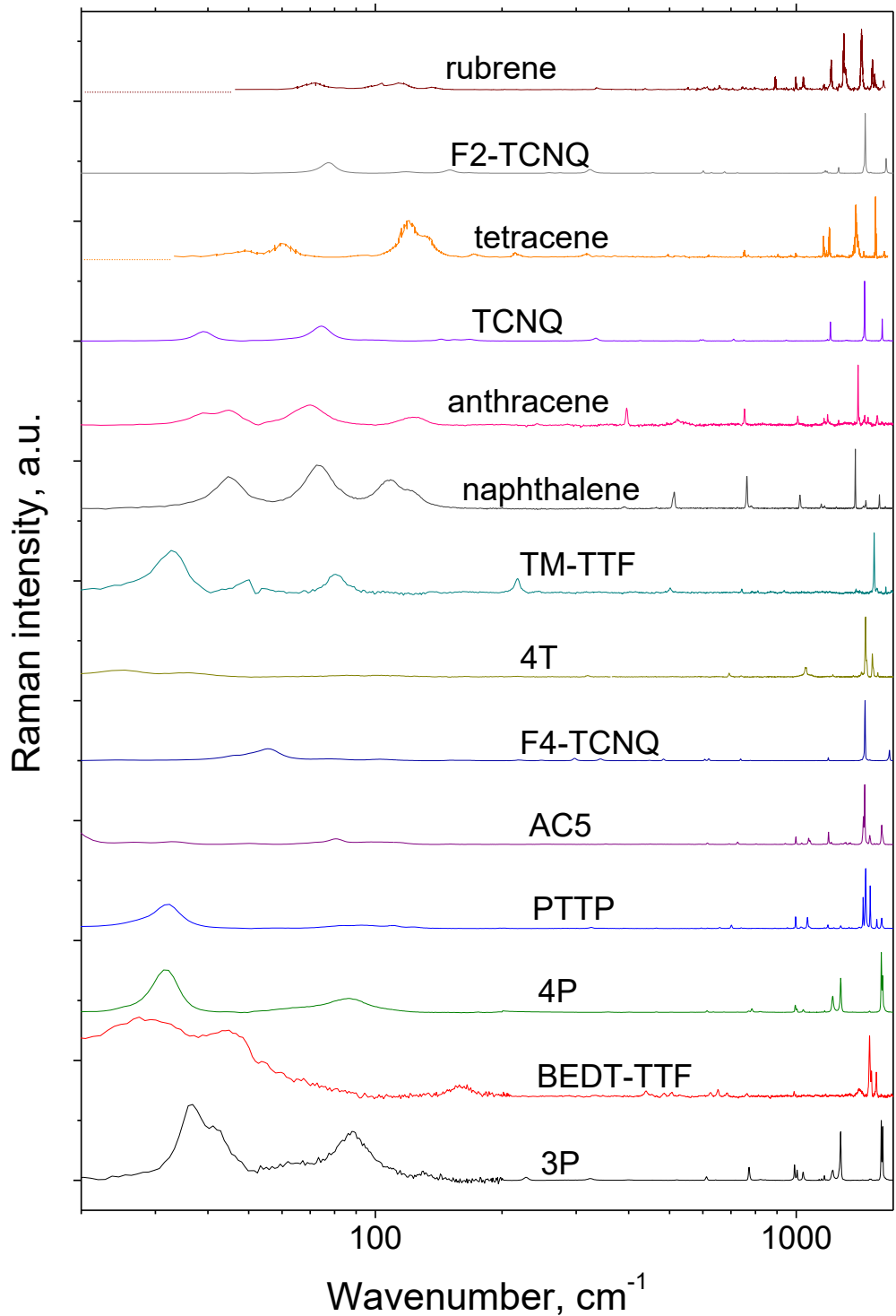

Figure S2. As-recorded Raman spectra of the investigated materials for the pump wavelength of 633 nm. Rubrene and tetracene spectra are from Ref. [10]. The spectra are normalized to the maximal intensity in the HF region and arranged according to the increase of the device charge mobility from bottom to top, see Table S1.

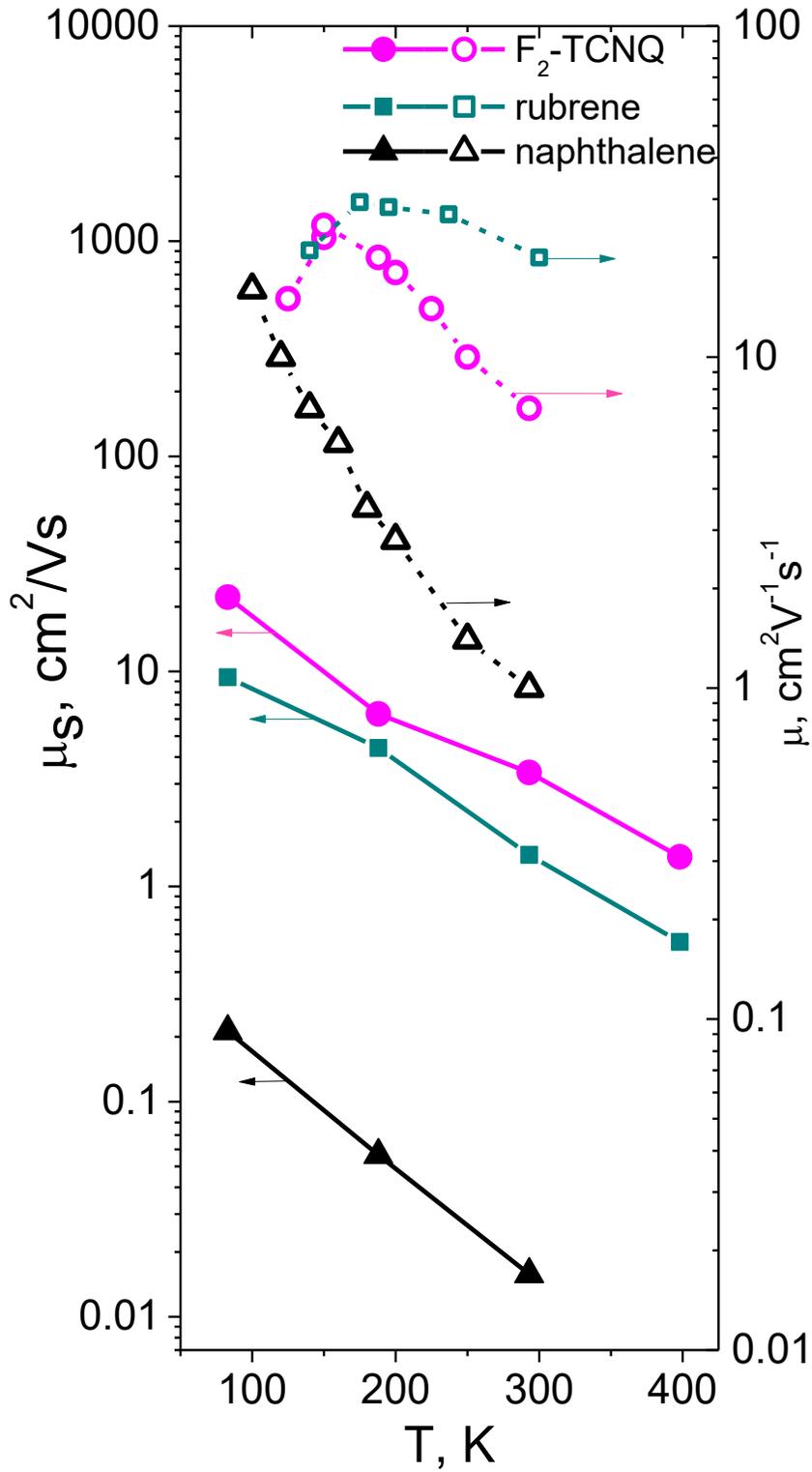

Figure S3. Correlation between the temperature dependencies of spectroscopic mobility, $\mu_s$, and device charge mobility, $\mu$, for OSs with band-like charge transport. Dependence of $J$ on $T$ is neglected.

## 2. 4. Device charge mobility data

| compound | Mobility type | Mobility value, $cm^2V^{-1}s^{-1}$ | Measurement type | Sample type | Reference |
|---|---|---|---|---|---|
| Rubrene | p | 13.5 | FET, air gap, linear | SC | [11] |
| F$_2$-TCNQ | n | 7 | FET, vacuum gap, linear | SC | [12] |
| Tetracene | p | 2.4 | FET, PDMS dielectric, saturation | SC | [13] |
| TCNQ | n | 1.6 | FET, vacuum gap, linear | SC | [11] |
| anthracene | p | 1.4 | ToF | SC | [14] |
| naphthalene | p | 1 | ToF | SC | [15] |
| TM-TTF | p | 0.68 | FET, polystyrene dielectric, linear | TF | [16] |
| 4T | P | 0.23 | FET, PDMS dielectric, saturation | SC | [17] |
| F$_4$-TCNQ | N | 0.2 | FET, air gap, linear | SC | [12] |
| PTTP | P | 0.064 | FET | SC | [18] |
| AC5 | P | 0.04 | FET | SC | [19] |
| 4P | P | 0.01 | FET | SC | [20] |
| BEDT-TTF | P | 0.0012 | FET | SC | [21] |
| 3P | P | 1E-4 | FET | TF | [22] |

Table S1. Device charge mobility data. FET = "field effect transistor", ToF= "time-of-flight", SC = "single crystal", TF = "thin film".